\documentclass{article}
\usepackage{arxiv}
\pdfoutput=1
\usepackage[utf8]{inputenc} 
\usepackage[T1]{fontenc}    
\usepackage{hyperref}       
\usepackage{url}            
\usepackage{booktabs}       
\usepackage{amsfonts}       
\usepackage{nicefrac}       
\usepackage{microtype}      
\usepackage{lipsum}

    \usepackage[compact]{titlesec}
    \titlespacing{\section}{0pt}{2ex}{1ex}
    \titlespacing{\subsection}{0pt}{1ex}{0ex}
     \titlespacing{\subsubsection}{0pt}{0.5ex}{0ex}

\author{
  Pratyush Garg \thanks{This research was supported under a grant from the Micron Foundation. This is a draft (as of the date noted above) of a paper that we are submitting for formal publication.} \\
  UCLA Electrical and Computer Engineering\\
  \texttt{pratyushg@ucla.edu} \\
   \And
  John Villasenor\\
  UCLA Electrical and Computer Engineering\\
  UCLA School of Law \\
  \texttt{villa@ee.ucla.edu} \\
  \And
   Virginia Foggo\\
  UCLA School of Law \\
  \texttt{vfoggo@ucla.edu} \\
  }
\usepackage[utf8]{inputenc}
\usepackage{amsmath}
\usepackage{amsmath}
\usepackage{url}
\usepackage{graphicx}  
\usepackage{float}  
\usepackage{url}
\usepackage{caption}
\usepackage{float}
\usepackage{graphicx}
\usepackage{amsmath}
\usepackage{nccmath}

\DeclareMathOperator{\EX}{\mathbb{E}}
\usepackage{array}
\newcolumntype{L}{>{\centering\arraybackslash}m{3cm}}
\usepackage[english]{isodate}
\begin{document}
\title{Fairness Metrics: A Comparative Analysis} 
\date{\small\today}
\maketitle
\begin{abstract}
Algorithmic fairness is receiving significant attention in the academic and broader literature due to the increasing use of predictive algorithms, including those based on artificial intelligence. One benefit of this trend is that algorithm designers and users have a growing set of fairness measures to choose from. However, this choice comes with the challenge of identifying how the different fairness measures relate to one another, as well as the extent to which they are compatible or mutually exclusive. We describe some of the most widely used fairness metrics using a common mathematical framework and present new results on the relationships among them. The results presented herein can help place both specialists and non-specialists in a better position to identify the metric best suited for their application and goals.
\end{abstract}


\section{Introduction}

The risks that algorithms---including those embedded in AI systems---can raise concerns relating to bias are well recognized. 
Addressing algorithmic bias requires an understanding of normative conceptions of fairness, and how they can be quantitatively measured. Moreover, algorithm designers and users will need to make informed decisions regarding which one or more of multiple possible fairness measures should be used for system assessments.
 
While fairness measures in relation to issues such as testing have been a topic of academic and broader interest for decades \cite{Hutchinson:2019:YTF:3287560.3287600}, recent years have seen rapidly growing interest among researchers within and beyond the technical community in the issue of \textit{algorithmic} fairness. The proliferation of papers describing fairness measures has spurred examination of the mathematical relationships among them.
For example, Kleinberg \textit{et al.}\@ \cite{1609.05807} considered three fairness measures (\textit{calibration within groups}, \textit{balance for the positive class}, and \textit{balance for the negative class}), and showed that ``except in highly constrained special cases, there is no method that can satisfy these three conditions simultaneously.'' Chouldechova \cite{1703.00056} has also written on the incompatibilities between fairness criteria
and has observed that \textit{test-fairness}, a measure ``originating in the field of educational and psychological testing'' can, when applied in the context of recidivism prediction, ``lead to considerable disparate impact when recidivism prevalence differs across groups.'' In another paper examining criminal risk assessments \cite{doi:10.1177/0049124118782533}, Berk \textit{et al.}\@ have observed that it is generally ``impossible to maximize accuracy and fairness at the same time, and impossible simultaneously to satisfy all kinds of fairness.''

Verma \textit{et al.}\@ \cite{Verma:2018:FDE:3194770.3194776} have ``collect[ed] the most prominent definitions of fairness for the algorithmic classification problem, explain[ed] the rationale behind these definitions, and demonstrate[d] each of them on a single unifying case-study.'' In a paper titled ``Fairness Through Awareness'' \cite{Dwork:2012:FTA:2090236.2090255}, Dwork \textit{et al.}\@, considered ``fairness in classification, where individuals are classified \ldots and the goal is to prevent discrimination against individuals based on their membership in some group, while maintaining utility for the classifier.'' Selbst \textit{et al.}\@ \cite{Selbst:2019:FAS:3287560.3287598}, have considered the broader societal context of seeking fair algorithms, and have argued that attempts to ``produce fairness-aware learning algorithms, and to intervene at different stages of a decision-making pipeline to produce `fair' outcomes'' can be  ``dangerously misguided when they enter the societal context that surrounds decision-making systems.''

While the papers cited above are drawn from technical publications, algorithmic fairness is also receiving rapidly growing attention in the legal scholarship. Examples include Mayson \cite{Mayson} who has argued that ``[a]lgorithmic risk assessment has revealed the inequality inherent in all prediction, forcing us to confront a problem much larger than the challenges of a new technology,'' MacCarthy \cite{MacCarthy}, who ``describe[d] and assesse[d] various group and individual statistical standards of fairness, including the mathematical conflict between the two that requires organizations to choose which measure to satisfy,'' and Hellman \cite{hellman}, who has explored the role of ``parity in the ratio of false positives to false negatives.''

Against this backdrop, the present paper offers several new contributions. First, as many previous publications each only consider a relatively small subset of fairness measures, differences in terminology and definitions can make comparisons among a broader group of measures difficult to perform. To address this, we describe some of the most widely cited fairness measures using a common mathematical and notational framework. Second and more substantively, we derive and discuss a set of mathematical relationships that facilitate comparisons among these metrics. The remainder of this paper is organized as follows. Section 2 presents metrics using a common mathematical framework and also aims to address some of the terminology variations across authors. Section 3 details the relationships between various metric pairs, including consideration of conditions under which they become mutually incompatible as well as trade-offs involved in selecting one metric over the other. Section 4 offers a discussion and conclusions.




\section{Fairness metrics}
 \subsection{Definitions}

    In the initial portion of the discussion herein we focus on binary predictions and their relationship to binary outcomes. We subsequently consider metrics that generate a continuous-valued score. In the binary context we assume that prediction involves a classifier that generates a binary prediction $\widetilde{y}$ based on a feature vector $x$ containing the data corresponding to a particular individual. If the distribution of $\widetilde{y}$ over $x$ is the same as that of the binary outcome $y$, we have perfect prediction. Hence, given a sufficiently large sample, the fairness and other attributes of the predictor can be evaluated by comparing predictions $\widetilde{y}$ with outcomes $y$. 
    
    We  denote individuals for whom $y=1$ as being members of the ``positive'' class and individuals for whom $y=0$ as being members of the ``negative'' class. For instance, if an algorithm predicts whether students will pass a test, $y$ will be 1 for the students who pass (and who are therefore members of the positive class) and 0 for students who fail (and are thus members of the negative class). While in this scenario the ``positive'' outcome $y=1$ corresponds to the desirable outcome, this will not always be the case. To take another example, in evaluating whether a parolee commits a new crime within a given time frame, $y=1$ can be used to denote the ``positive'' (but obviously undesirable) outcome that the parolee has committed a new crime.  
    
    We also assume that the dataset can be divided 
    into different groups based on attributes such as race, gender, etc., that may be of interest when evaluating whether an algorithm is biased. While there can be any number of such groups in different contexts, we will restrict our discussion to scenarios in which it is possible to identify two groups of individuals, with group membership indicated using a binary variable $G$. Finally, for some of the metrics we discuss, it will be necessary to consider
    a real-valued score $s$ $\in [0,1]$ that is computed for each individual which can optionally then be subject to thresholding to generate a binary prediction $\widetilde{y}$. 
In sum, we use the following notation:
    \begin{itemize}
    \itemsep-2pt
        \item $x$: feature vector 
        \item $G$: binary group index
        \item $\widetilde{y}$: binary prediction
        \item $y$: binary outcome
        \item $s$: classifier score 
        
    \end{itemize}
    
    We address some of the metrics that have received significant recent attention in the literature and/or that we believe present opportunities for more detailed comparative analysis. We do not claim to consider all possible fairness metrics (for example, we do not address \textit{fairness through awareness} as proposed by Dwork \textit{et al.}\@ \cite{Dwork:2012:FTA:2090236.2090255}). 
    
    To help clarify the discussion, in what follows we will sometimes refer to the following example involving loans. Consider two groups of people, which we will denote as the orange group (denoted group 0) and the blue group (group 1). The orange group has 60 members while the blue group has 40 members. If all the members of both groups were given loans, 40 members of the orange group and 20 members of the blue group would repay on time---these are the positive outcomes. However, among these 60 positive outcomes, we assume that the model predicted that only 28 of the orange group and 8 of the blue group will repay. These are the true positives for the two groups. Thus, the model correctly predicted the positive outcome of 36 individuals. 
    
    Similarly, if all members of both groups were given loans, 20 members of the orange group and 20 members of the blue group would default on the loan---these are the negative outcomes. From among these 40 negative outcomes, we assume that the model correctly predicted 12 of the defaulters in the orange group and 16 of the defaulters in the blue group. These are the true negatives for the two groups. In the aggregate, the model correctly predicted the negative outcome for 28 individuals. The table below summarizes these predictions and outcomes. We also add that, to make the illustration of the concepts more tractable, this example and the other numerical examples in this paper use small sample sizes (and in doing so implicitly assume statistical significance despite those small sample sizes). Of course, in a real scenario, to make reliable inferences on statistical outcomes and therefore on fairness assessments based on those outcomes, the sample sizes would need to be much larger. 
    
\begin{table}[H]
\centering
\caption{Illustrative Example}
\begin{tabular}{|c|L|L|L|}\hline
 & Orange Group & Blue Group & Total \\\hline
True positives  & 28  & 8  & 36\\\hline
False negatives  & 12 & 12 & 24 \\\hline
True negatives  & 12 & 16 & 28\\\hline
False positives  & 8 & 4 & 12 \\ \hline
Positives & 40 & 20 & 60 \\\hline
Negatives  & 20 & 20 & 40 \\ \hline
Total & 60 & 40  & 100\\\hline
\end{tabular}
\end{table}

    
    \subsection{Equalized odds and equality of opportunity}

    A predictor satisfies \textit{equalized odds} if both the true positive rate (TPR) and (separately) the false positive rate (FPR) are the same across groups.
    More formally, equalized odds requires that the group-specific TPR satisfy $p(\widetilde{y}=1|y=1,G=0) = p(\widetilde{y}=1|y=1,G=1)$ and that the group-specific FPR satisfy $p(\widetilde{y}=1|y=0,G=0) = p(\widetilde{y}=1|y=0,G=1)$. \cite{NIPS2016_6374} Since, the metric demands that the error rates be the same across groups, Chouldechova \cite{1703.00056}  describes this metric using the term ''error rate balance''. 
    In the example, for the orange and blue groups, the TPRs are $0.7$ and $0.4$ respectively, and the FPRs are $0.2$ and $0.4$ respectively. Thus, the example fails to satisfy equalized odds. 
    
    It is also worth noting that a related and less stringent metric, \textit{equality of opportunity} \cite{NIPS2016_6374}, can be defined by requiring only that the TPR be equal across groups, with no requirement imposed on the FPR. Thus, equalized odds implies equality of opportunity, though not vice versa. 
\\
    

    \subsection{Statistical parity}\label{sec:statparity}
    \textit{Statistical parity} \cite{1703.00056}, \cite{Verma:2018:FDE:3194770.3194776} (sometimes referred to as \textit{group fairness} \cite{Dwork:2012:FTA:2090236.2090255} or \textit{demographic parity} \cite{mehrabi2019survey}, \cite{NIPS2017_6995}) is achieved when members of both groups are predicted to belong to the positive class at the same rate. Mathematically, this means satisfying $p(\widetilde{y}=1|G=0)=p(\widetilde{y}=1|G=1)$. 
    Notably, this metric gives no consideration to the outcomes $y$. Therefore, when the base rates $p(y|G)$ differ across the two groups, statistical parity rules out the perfect predictor. 
    
    In the orange/blue group example, 36 of the 60 orange group members and 12 of the 40 blue group members were predicted to belong to the positive class. Since $p(\widetilde{y}=1|G=0)=36/60 = 0.6$ is not the same as $p(\widetilde{y}=1|G=1) =12/40 = 0.3$, the example does not satisfy statistical parity. 
    
    

    \subsection{Predictive parity}
    Consistent with Chouldechova \cite{1703.00056}, Verma \textit{et al.}\@ \cite{Verma:2018:FDE:3194770.3194776}, and MacCarthy \cite{MacCarthy} we consider that \textit{predictive parity} is satisfied when the positive predictive value(PPV) is the same for both groups. PPV is defined as the probability that individuals \textit{predicted} to belong to the positive class \textit{actually} belong to the positive class. Mathematically, predictive parity therefore requires $p(y=1|\widetilde{y}=1,G=0) = p(y=1|\widetilde{y}=1,G=1)$. 
    
    We note that some authors 
    define predictive parity in a more constrained manner, requiring not only parity for PPV, but also for its counterpart, negative predictive value (NPV), which requires additionally satisfying  $p(y=0|\widetilde{y}=0,G=0) = p(y=0|\widetilde{y}=0,G=1)$. Mayson \cite{Mayson} uses the term ``overall predictive parity'' to describe a predictor with equality across groups in both PPV and NPV, while Berk \textit{et al.}\@ \cite{doi:10.1177/0049124118782533} call this  ``conditional use accuracy equality''.
    
     
    
    In the loan example, the model predicted a total of 48 members across both groups to belong to the positive class. However, only 36 out of these were correct predictions; the rest were false positives. Hence, the overall PPV is $36/48=0.75$. Taken separately, the model predicted a total of 36 members of the orange group and a total of 12 members of the blue group to belong to the positive class. From these predictions, however, only 28 of the orange group and 8 of the blue group were correct. Therefore, the PPV for the orange group is $p(y=1|\widetilde{y}=1,G=0) =28/36 = 7/9 = 0.77$ and for the blue group is $p(y=1|\widetilde{y}=1,G=1)=8/12 = 0.66$. Since these values differ, the model from our example does not satisfy predictive parity.
    
    
    The fairness metrics discussed above can be evaluated using knowledge only of binary predictions and outcomes. By contrast, we now discuss a set of metrics involving explicit generation of a continuous-valued score $s$. Optionally, the score can serve as the input to a thresholding function that outputs a binary prediction, though scores can also be used directly, without any thresholding.

    \subsection{Calibration}
    
    An algorithm is \textit{calibrated} if for all scores $s$, the individuals who have the same score have the same probability of belonging to the positive class, regardless of group membership. \cite{1703.00056} \cite{Corbett-Davies:2017:ADM:3097983.3098095}  Mathematically, this is expressed through $p(y=1|S=s,G=0) = p(y=1|S=s,G=1)$.
    \ This metric has been termed  \textit{test-fairness} by Chouldechova \cite{1703.00056}, Verma \textit{et al}.\@  \cite{Verma:2018:FDE:3194770.3194776}, and Mehrabi \textit{et al.}\@ \cite{mehrabi2019survey} and as \textit{matching conditional frequencies} by Hardt \textit{et al.}\@ \cite{NIPS2016_6374}
    

    There is another related metric termed \textit{well-calibration} \cite{Verma:2018:FDE:3194770.3194776} or \textit{calibration within groups} \cite{1609.05807}\cite{doi:10.1177/0049124118782533} that imposes an additional, more stringent condition. In order for a model to be well-calibrated (or to have calibration within groups), individuals assigned score \textit{s} must have probability $s$ of belonging to the positive class. 
    If this condition is satisfied, then test-fairness will also automatically be satisfied, though the reverse does not hold.
       The difference between calibration and well-calibration is simply one of mapping; the scores of a calibrated predictor can, using a suitable transformation, be converted to scores satisfying well-calibration.

    \subsection{Balance for positive/negative class} \label{balance}
    Kleinberg \textit{et al.}\@ \cite{1609.05807} have noted that when the average score \textit{s} for all individuals constituting the group-specific positive class is the same for both groups of interest, it can be said that there exists \textit{balance for the positive class}. Similarly, \textit{balance for the negative class} is satisfied when the average score \textit{s} for members of the negative class are equal, regardless of group membership. Mathematically this is expressed in terms of expected values. For the negative class, balance requires $\EX[s|y=0,G=0] = \EX[s|y=0,G=1]$, and for the positive class balance requires $\EX[s|y=1,G=0] = \EX[s|y=1,G=1]$. 
    
    As Pleiss \textit{et al.}\@ \cite{Pleiss:2017:FC:3295222.3295319} have explained, this can be viewed as a generalization of the equalized odds metric to non-binary cases. To see this, note that when the score $s$ can take on only the two values 0 and 1, the score itself is the prediction and the term $\EX[s|y=0,G]$ then represents the false positive rate and the term $\EX[s|y=1,G]$ represents the true positive rate. And, as noted above, under equalized odds the TPR and FPR are equal across groups.

\section{Comparisons and trade-offs between metrics}


Different fairness metrics formalize varying intuitive notions of fairness. This raises the question of the conditions under which more than one metric can be simultaneously satisfied, and relatedly, the ways in which different metrics might be in tension. 

We will analyze metrics under the assumption that the ``base rate'' differs across groups. The base rate of a group is the ratio of people in the group who belong to the positive class ($y=1$) to the total number of people in that group. Thus, having non-equal base rates across groups means that $p(y=1|G=0) \neq p(y=1|G=1)$. 

In the subsequent discussion, to simplify the equations we will use the following terms as defined here:

\begin{itemize}
    \item $\text{TPR}_g=p(\widetilde{y}=1|y=1,G=g), g \in \{0,1\}$ --- the group-specific true positive rates. 
    \item  $\text{FPR}_g=p(\widetilde{y}=1|y=0,G=g), g \in \{0,1\}$ --- the group-specific false positive rates.
    \item  $\text{PPV}_g=p(y=1|\widetilde{y}=1,G=g), g \in \{0,1\}$ --- the group-specific positive predictive value. 
\end{itemize}





\subsection{Statistical parity, equalized odds and predictive parity}
Trade-offs among statistical parity, equalized odds and predictive parity have received significant attention in algorithmic fairness literature under a variety of formulations. Chouldechova \cite{1703.00056} articulates the tradeoffs between predictive parity and equalized odds empirically. Kleinberg \textit{et al.}\@ \cite{1609.05807}, while not directly referring to these terms, give a closely related result, writing that ``the calibration condition and the balance conditions for the positive and negative classes" are ``in general incompatible with each other; they can only be simultaneously satisfied in certain highly constrained cases.'' Berk\textit{ et al.}\@ \cite{doi:10.1177/0049124118782533} discuss incompatibility among metrics by taking several cases and examining the trade-offs in those scenarios. 

We offer a common mathematical framework to examine trade-offs among  statistical parity, equalized odds and predictive parity. We provide proofs regarding the combination of all three of these metrics and also explore conditions under which it may be possible to simultaneously satisfy two metrics. To provide an initial framing, it is interesting to note that using the basic probability relation $p(A,B) = p(A|B)p(B) = p(B|A)p(A)$, the respective probability distributions associated with each of these three metrics can be expressed as follows:

\begin{flalign} \label{eq:1}
\begin{aligned}
    p(y,\widetilde{y}|G) = \underbrace{p(y|\widetilde{y},G)}_{\text{Predictive Parity}} \ \times \ \underbrace{p(\widetilde{y}|G)}_{\text{Statistical Parity}} = \underbrace{p(\widetilde{y}|y,G)}_{\text{Equalized Odds}} \ \times \ \ \underbrace{p(y|G)}_{\text{Base Rate}}
\end{aligned}
\end{flalign}


\subsubsection{All three?}

This section considers the feasibility of satisfying all three metrics under the assumption of unequal base rates. We start by assuming a predictor that satisfies both statistical parity and equalized odds, and then examining if it can also satisfy predictive parity. From equation \ref{eq:1}, we have:
\begin{flalign} \label{eq:2}
\begin{aligned}
p(y=1|\widetilde{y}=1,G) = \frac{p(\widetilde{y}=1|y=1,G) \times p(y=1|G)}{p(\widetilde{y}=1|G)}
\end{aligned}
\end{flalign}
Since the predictor satisfies equalized odds, the TPR must be the same across groups and therefore, we denote $\text{TPR}_0 = \text{TPR}_1 = \text{TPR}$.
And since the predictor by definition also satisfies statistical parity,     $p(\widetilde{y}=1|G=0)=p(\widetilde{y}=1|G=1)=p(\widetilde{y}=1)$ Imposing these conditions and taking the difference of the PPV values of the two groups gives:
\begin{flalign} \label{eq:3}
\begin{aligned}
    p(y=1|\widetilde{y}=1,G=0) - p&(y=1|\widetilde{y}=1,G=1) =  \frac{\text{TPR}[p(y=1|G=0) - p(y=1|G=1)]}{p(\widetilde{y}=1)}
\end{aligned}
\end{flalign}
Predictive parity requires that the PPV be equal across both groups, and therefore that the difference on the left side of the above equation be zero, which in turn can only occur when the base rates across the two groups are equal as indicated by the right side of the equation. Note that equalized odds requires \textit{both} the TPR and the FPR to be the same. However, by demonstrating the incompatibility of just TPR in this section, we provide a sufficient proof for the incompatibility of equalized odds in the given scenario. 
Thus, when the two groups have unequal base rates, satisfying all three of statistical parity, predictive parity and equalized odds is impossible. This remains true even if the predictor is perfect, as a perfect predictor can not (when the base rates are unequal) satisfy statistical parity. 

\subsubsection{Statistical parity and predictive parity}
We now consider conditions under which a predictor can satisfy both statistical and predictive parity. 
Recall that when statistical parity holds, we have $p(\widetilde{y}=1|G=0) = p(\widetilde{y}=1|G=1)=p(\widetilde{y}=1)$. 
 Taking the difference in PPV across the two groups gives:
\begin{flalign}\label{eq:4}
\begin{aligned}
    p(y=1|\widetilde{y}=1,G=0) - p&(y=1|\widetilde{y}=1,G=1) = \frac{\text{TPR}_0p(y=1|G=0) - \text{TPR}_1p(y=1|G=1)}{p(\widetilde{y}=1)}
\end{aligned}
\end{flalign}
Under predictive parity the left side of the equation must be zero, which in turn requires that 
the ratio of the true positive rates of the two groups 
be the reciprocal of the ratio of the base rates, i.e.:

\begin{flalign}\label{eq:5}
\begin{aligned}
   \frac{\text{TPR}_0}{\text{TPR}_1}  =  \frac{p(\widetilde{y}=1|y=1,G=0)}{p(\widetilde{y}=1|y=1,G=1)} = \frac{p(y=1|G=1)}{p(y=1|G=0)} = \frac{\text{Base Rate of Group 1}}{\text{Base Rate of Group 0}}
\end{aligned}
\end{flalign}
Thus, while statistical and predictive parity can be simultaneously satisfied even with different base rates, the utility of such a predictor is limited when the ratio of the base rates differs significantly from 1, as this forces the true positive rate for one of the groups to be very low.


As mentioned above, the definition of predictive parity used here, consistent with \cite{1703.00056}, only requires different groups to have the same PPV. However, if we were to consider the ``overall predictive parity'' \cite{Mayson}, and require a predictor to also have the same NPV across groups, the system would be overconstrained and it would not generally be possible to simultaneously satisfy statistical parity and predictive parity. 

\subsubsection{Equalized odds and predictive parity}
Chouldechova (2017) \cite{1703.00056} observes that "predictive parity is incompatible with error rate balance when prevalence differs across groups," (p. 5). (As noted above, Chouldechova uses ``error rate balance'' to describe what we refer to here as equalized odds.) We explore this incompatibility 
in more detail. As before when equalized odds and predictive parity are satisfied, we have $\text{TPR}_0=\text{TPR}_1$, $\text{FPR}_0=\text{FPR}_1$, and $\text{PPV}_0=\text{PPV}_1$.  Noting that 
\[p(\widetilde{y}=1|G) = \sum_y p(\widetilde{y}=1|y,G) p(y|G) =  p(\widetilde{y}=1|y=1,G)p(y=1|G) + p(\widetilde{y}=1|y=0,G)p(y=0|G)\] 
\begin{flalign}\label{eq:6}
\begin{aligned}
    \implies p(\widetilde{y}=1|G) = \text{TPR}_0p(y=1|G) + \text{FPR}_0p(y=0|G)
\end{aligned}
\end{flalign}
and considering equation \ref{eq:1}, we can write:

\[p(\widetilde{y}=1|y=1,G=0)p(y=1|G=0) = p(y=1|\widetilde{y}=1,G=0) [\text{TPR}_0 p(y=1|G=0) + \text{FPR}_0p(y=0|G=0)]\]
\[  \implies \text{TPR}_0p(y=1|G=0) = \text{PPV}_0 [\text{TPR}_0 p(y=1|G=0) + \text{FPR}_0p(y=0|G=0)] \]
\[ \text{TPR}_0 p(y=1|G=0) = \text{PPV}_0 [\text{TPR}_0 p(y=1|G=0) + \text{FPR}_0(1-p(y=1|G=0))] \]
\begin{flalign}\label{eq:7}
\begin{aligned}
\implies  p(y=1|G=0) = \frac{\text{PPV}_0\text{FPR}_0}{\text{PPV}_0\text{FPR}_0 + (1-\text{PPV}_0)\text{TPR}_0} 
\end{aligned}
\end{flalign}
\begin{flalign}\label{eq:11}
\begin{aligned}
 \text{Likewise, }p(y=1|G=1) = \frac{\text{PPV}_1\text{FPR}_1}{\text{PPV}_1\text{FPR}_1 + (1-\text{PPV}_1)\text{TPR}_1}
\end{aligned}
\end{flalign}

But since $\text{TPR}_0=\text{TPR}_1$, $\text{FPR}_0=\text{FPR}_1$ and $\text{PPV}_0=\text{PPV}_1$, equations 7 and 8 will be identical, so base rates for groups 1 and 2 will be the same: $p(y=1|G=0) = p(y=1|G=1)$. Hence, in the absence of perfect prediction, the base rates have to be equal for both equalized odds and predictive parity to simultaneously hold. When perfect prediction is achieved, equations \ref{eq:7} and \ref{eq:11} take on the indefinite form $0/0$ so therefore do not convey anything definitive about base rates in that scenario. 

We also note that the metric equal opportunity (a less strict counterpart to equalized odds that requires only equal TPR across groups) is compatible with predictive parity. This is evident from equations \ref{eq:7} and \ref{eq:11} when the condition $\text{FPR}_0=\text{FPR}_1$ is removed, thereby allowing equalized opportunity and predictive parity to be simultaneously satisfied even with unequal base rates. However, achieving this condition with unequal base rates will require that the FPR differs across the groups. When the difference between the base rates is large, the variation between group-specific FPRs may have to be significant which may reduce suitability for some applications. Hence, while equal opportunity and predictive parity are compatible in the presence of unequal base rates, practitioners should consider the cost (in terms of FPR difference) before attempting to simultaneously achieve both. A similar analysis is possible when we considering parity in negative predictive value instead of positive predictive value, i.e. equal opportunity and parity in NPV are compatible, but only at the cost of variation between group-specific true negative rates (TNRs).


\subsubsection{Equalized odds and statistical parity}
We now consider if a predictor can simultaneously satisfy equalized odds and statistical parity. As before, for $\text{TPR} = \text{TPR}_0 = \text{TPR}_1$ (equal TPR) and $\text{FPR} = \text{FPR}_0 =\text{FPR}_1$ (equal FPR):
\[ p(\widetilde{y}=1|G)= \text{TPR}[p(y=1|G)] + \text{FPR}[p(y=0|G)]\\ \]

\begin{flalign}\label{eq:8}
\begin{aligned}
\implies  p(\widetilde{y}=1|G=0)-p(\widetilde{y}=1|G=1) = &\text{TPR}[p(y=1|G=0)-p(y=1|G=1)] \\ +& \text{FPR}[p(y=0|G=0)-p(y=0|G=1)]\\ 
\end{aligned}
\end{flalign}
\begin{flalign}\label{eq:8}
\begin{aligned}
 \implies  p(\widetilde{y}=1|G=0)-p(\widetilde{y}=1|G=1) = (\text{TPR}-\text{FPR})[p(y=1|G=0)-p(y=1|G=1)]\\
\end{aligned}
\end{flalign}

Statistical parity requires the left side of equation \ref{eq:8} to be zero. For the equation to hold, this means the right side must also be zero, which can only occur when either $\text{TPR}=\text{FPR}$ or $p(y=1|G=0)=p(y=1|G=1)$. The latter case, however, violates the assumption that the base rates are different. Therefore to have both statistical parity and equalized odds, the only possibility is to have $\text{TPR}=\text{FPR}$, i.e. the false positive rate and the true positive rate have to be equal. Thus, while simultaneously achieving statistical parity and equalized odds is mathematically possible, it is not particularly useful since the goal is typically to develop a predictor in which the TPR is significantly higher than the FPR.




\subsection{Predictive parity and calibration}

There has been some confusion in the literature regarding the relationship between these two metrics. Chouldechova \cite{1703.00056} has correctly mentioned that ``While predictive parity and calibration look like very similar criteria, well-calibrated scores can fail to satisfy predictive parity at a given threshold,'' However, in other papers, the discussion on these two metrics sometimes gives less attention to the specifics of how they relate than we think is merited.
Given this backdrop, in the present section we explain the difference between predictive parity and calibration 
with a mathematical derivation and an example. 

It is possible to view calibration as a generalization of predictive parity to the non-binary setting. 
 The score $s$ discussed in relation to calibration is generally continuous-valued. However, in the special case in which it is limited to the two values 0 and 1 and therefore becomes the prediction itself, achieving calibration is the same as achieving equality across groups in both PPV and NPV. Thus, this satisfies both predictive parity (due to equality across groups in PPV) as well as ''overall predictive parity'' (due to equality across groups in both PPV and NPV).

Of course, in general the score $s$ is not binary. A continuous-valued score can be binarized through a thresholding operation to generate
a binary prediction $\widetilde{y}$. However, it is \textit{not} the case that thresholding a calibrated score in this manner necessarily leads to predictive parity.

To prove this, consider a threshold $s_{th} \in [0,1]$, such that $\forall s>s_{th}, \widetilde{y}=1$ and $\widetilde{y}=0$ otherwise. Hence, the distribution relevant to  predictive parity $p(y=1|\widetilde{y}=1,G)$ can be expressed
$p(y=1|s>s_{th},G)$. Using this we can write:

\begin{flalign}
\begin{aligned}
    p(y,s>s_{th}|G)  = \int_{s_{th}}^1 \underbrace{p(y|s,G)}_{\text{calibration term}}p(s|G)ds
\end{aligned}
\end{flalign}
\begin{flalign}\label{eqn:12}
\begin{aligned}
   \implies \underbrace{p(y|s>s_{th},G)}_{\text{predictive parity term}} = \frac{\int_{s_{th}}^1 p(y|s,G)p(s|G)ds}{\int_{s_{th}}^1 p(s|G)ds}
\end{aligned}
\end{flalign}
The above equation relates predictive parity to calibration, showing that even 
when the calibration term $p(y|s,G)$ is the same for both groups, the probability distribution of the score, expressed in equation \ref{eqn:12} through $p(s|G)$, can vary across groups in a way that causes predictive parity not to be satisfied.
To make this more intuitive, we will consider a special case where there are only two score values $s_1$ and $s_2$ above the threshold $s_{th}$ such that $p(s|G) \neq 0$. In other words, all individuals who receive risk scores above the threshold have the possibility of receiving one of only two scores, $s_1$ or $s_2$. Hence, $p(s>s_{th}|G) = p(s=s_1|G)+p(s=s_2|G)$. 

Under this special case equation \ref{eqn:12} reduces to:

\begin{flalign}\label{eqn:13}
\begin{aligned}
    p(y=1|\widetilde{y}=1,G) = \frac{p(y=1|s=s_1, G)p(s=s_1|G) + p(y=1|s=s_2, G)p(s=s_2|G)}{p(s=s_1|G)+p(s=s_2|G)}
\end{aligned}
\end{flalign}

Using this scenario, consider an example in which we have 100 people in each of two groups: orange and blue (this is a new example, unrelated to the example using orange and blue groups introduced earlier in the paper). Consider further an algorithm that only gives one of three possible scores (0.25, 0.5 or 0.75) to every individual's loan application. Suppose that scores are being binarized using a threshold of 0.49, such that any individual with a score above 0.49 is deemed to belong to the positive class. In this example, this would mean there are two possible scores (0.5 and 0.75) that can lead to a positive prediction. This is illustrated in Table \ref{tab:ppvscal} given below.

\begin{table}[H]
\centering
\caption{Predictive Parity and Calibration Example}
\begin{tabular}{|c|L|L|L|}\hline
Score & Orange Group & Blue Group & Prediction after threshold with $s_{th}=0.49$ \\\hline
0.25 & 40 (16) & 40 (16) & Negative\\\hline
0.5 & 20 (10) & 40 (20) & Positive\\\hline
0.75 & 40 (30) & 20 (15) & Positive \\\hline
Total & 100(56) & 100(51) & \\\hline
\end{tabular}
\label{tab:ppvscal}
\end{table}

The first column represents the score that the model assigned. In the second and third columns, the numbers outside the parentheses convey the number of people in the group assigned that score. The numbers in parentheses represent the number of people from those assigned that score who actually belong to the positive class.
In this example the predictor is calibrated, since given a score, the fraction of people who actually belong to the positive class is independent of the group. For example, for score 0.5, $10/20=0.5=50\%$ of the people in the orange group with that score and $20/40=0.5=50\%$ of the people in the blue group with that score belong to the positive class. 

Does this model satisfy predictive parity? Choosing 0.49 as the threshold gives a total of 60 positive predictions for both the orange group and the blue group. However, of the people with scores greater than 0.49, only 40 members in the orange group and 35 members of the blue group are actually in the positive class, resulting in a PPV of $40/60=0.66$ for the orange group and $35/60=0.583$ for the blue group. Thus, while the predictor is calibrated, choosing a threshold of 0.49 does not lead to a set of binary predictions that satisfy predictive parity. 




It is also interesting to note that if all persons who had a score of 0.25 are instead given a score of 0.4, the model will not only be calibrated (because, as before, people with the same score have the same probability of belonging to the positive class) but also \textit{well}-calibrated (because, due to this change in scoring, for all scores the score itself would give the probability of belonging to the positive class). However, this change in score would have no impact on the thresholding example above, illustrating that even a well-calibrated model does not, after applying a threshold to produce binary predictions, necessarily satisfy predictive parity.


This discussion can be further generalized. Consider the comparison between statistical parity and calibration given by Kleinberg \textit{et al.}\@ \cite{1609.05807}. Kleinberg \textit{et al.}\@ give an alternate definition of statistical parity which is independent of the choice of threshold, defining it as the condition in which both groups have the same average score.
They then prove that when there are unequal base rates and imperfect prediction, statistical parity is incompatible with well-calibration.  

In one sense, the Kleinberg \textit{et al.}\@ definition of statistical parity can be understood as a generalization of the binary case. This is because in the special case where the score itself is binary and  represents the prediction, the Kleinberg \textit{et al.}\@ definition reduces to the common  (i.e., binary) definition of statistical parity that we have provided in section \ref{sec:statparity}. However, when there is \textit{both} a score $s$ \textit{and} a binary prediction $\widetilde{y}$ obtained by thresholding the score, the fact that there was statistical parity prior to thresholding in accordance with the definition from Kleinberg \textit{et al.}\@ does not necessarily imply that there will be statistical parity of the binary predictions in accordance with \ref{sec:statparity}. Stated another way, if, prior to thresholding, the average score $s$ is the same across two groups, it does not necessarily follow that thresholding will produce predictions $\widetilde{y}$ in which $p(\widetilde{y}|G=0)=p(\widetilde{y}|G=1)$.


Under the standard definition of statistical parity (in \ref{sec:statparity})
it is possible to simultaneously satisfy both statistical parity and calibration. 
Consider, again, the example in table \ref{tab:ppvscal}, where despite different base rates, when a threshold of 0.49 is applied to binarize the calibrated scores, it also satisfies statistical parity
(for both groups of 100, 60 individuals are predicted to belong to the positive class). The base rates are also different since a total of 56 people from the orange group and 51 people from the blue group belong to the positive class. Again, as before, if all persons who had a score of 0.25 were instead given a score of 0.4, the model will be well-calibrated and still satisfy statistical parity for the threshold of 0.49. Also, the prediction is clearly imperfect. Thus, statistical parity (as we have defined it in section \ref{sec:statparity} above) is not necessarily incompatible with having calibration (and well-calibration) even when the base rates are different and there is imperfect prediction. 

This underscores the importance of being attentive to the difference between fairness metrics designed for use in relation to scores $s$ and those intended for use with binary predictions. While it is straightforward to convert scores to binary values through thresholding, the ease of the conversion masks important complexities regarding the extent to which fairness metrics might be met after the thresholding process. In the example above, the scores were calibrated, and upon application of a threshold of 0.49, the binary predictions exhibited statistical parity. However, the same underlying distribution of scores, had they been subject to a threshold of 0.55, would have yielded predictions \textit{without} statistical parity. More 
generally, we believe that there is room for---and a need for---fairness metrics in relation to both continuous scores $s$ as well as binary predictions $\widetilde{y}$. But the benefits offered by having a greater number of tools for examining fairness must be balanced with an awareness of the issues that can arise when working across these two domains.

\section{Discussion and conclusions}

Several conclusions arise from the discussion above. With respect to metrics such as statistical parity, equalized odds and predictive parity that evaluate fairness by comparing binary predictions with binary outcomes, pairwise combinations of metrics can be simultaneously satisfied only under very limited conditions, if at all. For example, when the base rates are different, satisfying both statistical parity and predictive parity requires that the ratio of group-specific true positive rates be the inverse of the ratio of the base rates. When the ratio of the base rates does not deviate too far from 1, this constraint can be met while also preserving high true positive rates for both groups. The question of  how far a deviation from 1 in the base rate ratio (and therefore in the inverse of the TPR ratio) would still be acceptable would of course be context dependent. For large (or small) base rate ratios, the resulting predictor would necessarily have a low true positive rate for one of the groups. We also showed that equalized odds and predictive parity are incompatible when the base rates across two groups differs. In addition, we showed that, given unequal base rates, equalized odds and statistical parity can only be simultaneously met in the rather impractical case where the true positive and false positive rates are equal. The above results illustrate that, for unequal base rates, a given pair of metrics can be 1) mathematically incompatible, 2) mathematically compatible but under constraints that are problematic from a policy standpoint, or 3) mathematically compatible under constraints that are consistent with positive policy outcomes.

With respect to more generalized predictors that generate a (non-binary) score, we explored the relationship between calibration (computed with respect to score that can in general be continuous) and predictive parity when that score is subject to thresholding to generate a binary prediction. We observed the value of utilizing fairness metrics designed for use in relation to scores, while also emphasizing some of the complexities involved in working across continuous and binary domains. In particular, we noted that calibrated scores, after thresholding, may still generate predictions that satisfy statistical parity, though threshold choice can play an important role in determining whether or not such conditions are satisfied. 

In closing, we note that in addition to the above considerations, fairness metric selection can also be guided by the observability of each statistic in practice.  For example, in the context of loan approvals, loans will typically only be given to the subset of applicants who are predicted to repay. When this occurs, it will be impossible to observe statistics such as false negatives, true negatives, negative predictive value, true positive rate, and true negative rate. By contrast, the positive predictive value will be readily observable, suggesting that a metric such as predictive parity would be easier to evaluate in practice.

\bibliographystyle{ieeetran}
\bibliography{refer.bib}

\end{document}